\documentclass[english,twocolumn,times,amsmath,amssymb,pra, showpacs]{revtex4-1}
\usepackage{graphicx,epsfig,epstopdf}
\usepackage{mathrsfs}
\usepackage{dcolumn}
\usepackage{bm}
\usepackage{color}

\begin{document}

\title{Nonlinear light scattering in molecules triggered by an impulsive X-ray Raman process}
\author{Konstantin E. Dorfman}
\email{Email: kdorfman@uci.edu}
\author{Kochise Bennett}
\email{Email: kcbennet@uci.edu}
\author{Yu Zhang}
\author{Shaul Mukamel}
\affiliation{University of California, Irvine, California 92697-2025}
\date{\today}

\begin{abstract}

The time-and-frequency resolved nonlinear light scattering (NLS) signals from a time evolving charge distribution of valence electrons prepared by impulsive X-ray pulses are calculated using a superoperator Green's function formalism. The signal consists of a coherent $\sim N^2$-scaling difference frequency generation and an incoherent fluorescence $\sim N$-scaling component where $N$ is the number of active molecules. The former is given by the classical Larmor formula based on the time-dependent charge density.  The latter requires additional information about the electronic structure and may be recast in terms of  transition amplitudes representing quantum matter pathways.
\end{abstract}

\maketitle

\section{Introduction}

Newly developed attosecond X-ray sources can excite molecular electronic states impulsively, paving the way for novel spectroscopic probes of electronic structure and correlations \cite{ull12, gal12, emm10, kap07, kra09, wor11, hen01, pop12, zho10,biggs13}. 
Thanks to their broad bandwidth, such pulses can prepare molecules or molecular ions in non-stationary superpositions of electronic states \cite{Ame11}. These states can be probed by various nonlinear optical techniques, such as photoelectron spectroscopy \cite{santra,cederbaum}.

Traditional X-ray experiments \cite{gro08}, including X-ray absorption near-edge structure (XANES), resonant and nonresonant X-ray emission spectroscopy (RXES and NXES), Auger electron spectroscopy (AES), and X-ray diffraction \cite{Dix12} (XD), are mostly related to single-particle characteristics of the many-body ground-state wave function. Quantum coherence (i.e., the phase in a superposition of states), which is accessible by nonlinear spectroscopy, does not play a role in these techniques.  Coherent X-ray sources can look into these quantum effects. In particular, the large bandwidth (10 eV for an 100 as pulse) can be used to create electronic coherences which provide a higher level of information about orbitals that goes beyond the charge density \cite{bre05}. Nonlinear effects in the X-ray regime have long been observed in the frequency domain \cite{ada03}, including parametric downconversion \cite{eis71,yod98, ada00} hard X-ray frequency doubling \cite{naz03} and two-photon X-ray fluorescence \cite{ban82}. Combined optical/X-ray nonlinear techniques can monitor excitations of optically excited states \cite{muk10, biggs, ZBHG12, eis71}.  All-X-ray nonlinearities such as pump-probe have been demonstrated as well \cite{naz03, pop10, Ame11, gelmukhanov, yod98, ada00, har09}.

Classically, spontaneous emission is related to the acceleration of charges \cite{cederbaum}. In the semiclassical approach (classical field/quantum matter) \cite{scully} the emission spectrum is thus calculated from a time dependent charge density that acts as a source. No other information about the matter is needed. This results in the Larmor formula for the power emitted by a radiating charge (Eq. \ref{eq:scoht}) \cite{jackson, cederbaum}.  In a more rigorous description where both field and matter are treated quantum mechanically \cite{Tan01} we must work in the joint field/matter space and calculate the signal perturbatively in the radiation-matter coupling. Radiation back-reaction is then included implicitly, through the joint dynamics of field and matter.

\begin{figure}[t]
\begin{center}
\includegraphics[trim=0cm 0cm 0cm 0cm,angle=0, width=0.48\textwidth]{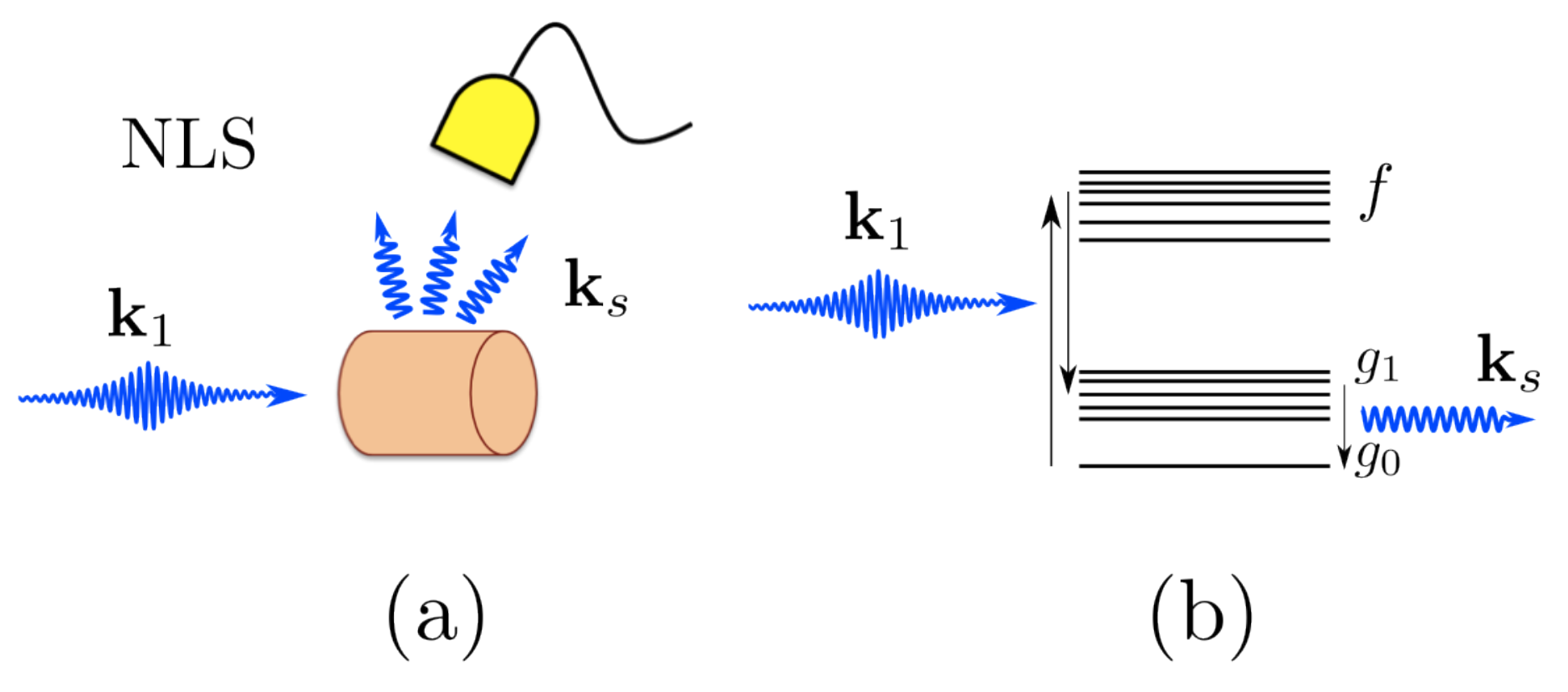}
\end{center}
\caption{(Color online) (a) - setup for the X-ray induced NLS, (b) - level scheme of the model system, $g_0$,..., $g_1$ represent valence states, $f$ are core excited states.}
\label{fig:set}
\end{figure}

In the present work we study the nonlinear light scattering (NLS) caused by the time evolving superposition of electronic excitations. In order to create such superpositions the coherent X-ray source with broad ($\sim$10 eV) bandwidth is required. We use the many-body Green's function formalism \cite{Har08} to study NLS corresponding to transitions between valence excitations triggered by stimulated X-ray Raman process \cite{biggs13,SDCF94,Niri} (see Fig. \ref{fig:set}). Note, that NLS results from the single pulse used for excitation; there is no probe pulse to simulate the emission. The resulting time and/or frequency gated signals \cite{Dor12,Sto94} provides direct information on valence electron motions, coherences and correlations \cite{Salam,biggs13}. 
We identify two possible mechanisms for this process to quadratic order in X-ray pulse intensity: difference frequency generation (DFG) and stimulated Raman induced fluorescence (SRIF). The former is a coherent process associated with long range  valence state coherence, which can be described in terms of the time-dependent charge density created by the Raman process i.e. the classical Larmor formula. It is highly directional (phase matched) and scales as $N(N-1)$ with the number of active molecules. The latter is an incoherent process coming from excited state populations which requires more detailed information about the excited states than the simple charge density. It produces an isotropic signal and scales as $N$. For small samples, the coherent process is dominated by a short range coherence coming from the neighbor molecules that gives rise to an isotropic emission and scales linearly with $N$. As an example we compare both contributions for cysteine excited at the S, O and N core transitions.

The paper is organized as follows. In Section II we present general formal expressions for the coherent and incoherent signals that allow to incorporate an arbitrary sequence of preparation pulses. The coherent signal is given by the time dependent polarization whereas the incoherent is expressed in terms of transition amplitude.
In Section III we skip the preparation stage (the shaded area in Fig. 2). We assume that the system has been prepared in a superposition state and calculate the spontaneous light emission.
In Section IV we apply the results of Section III to compute the spectra of cysteine induced by a stimulated X-ray Raman process. We further show that when the preparation is included as in Section II the coherent and incoherent signals represent difference frequency generation and fluorescence, respectively.

\section{Coherent vs Incoherent spontaneous light scattering signals}

\begin{figure}[t]
\begin{center}
\includegraphics[trim=0cm 0cm 0cm 0cm,angle=0, width=0.48\textwidth]{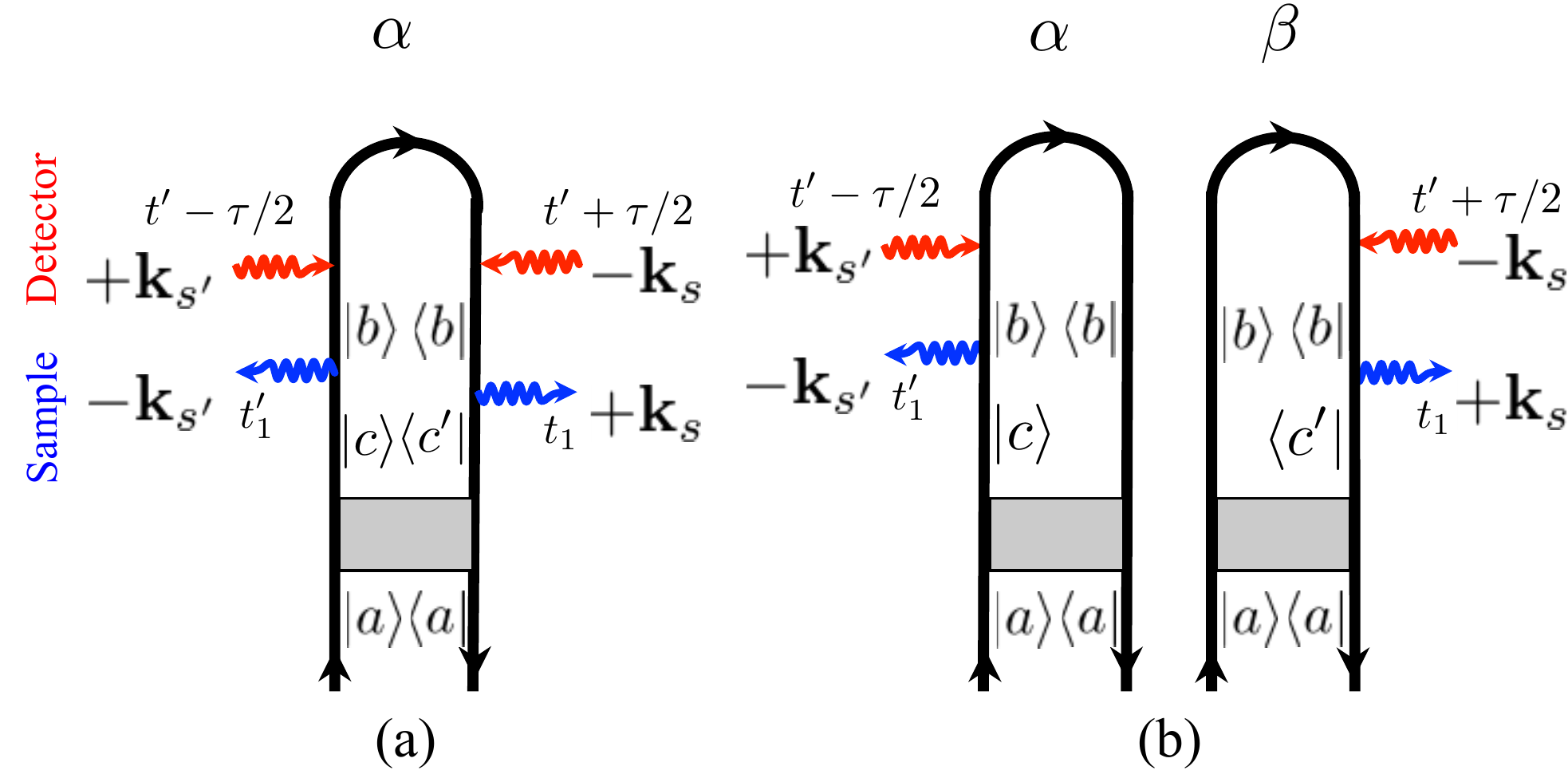}
\end{center}
\caption{(Color online) ($a$) loop diagram for the bare incoherent signal in a gated measurement caused by a single molecule $\alpha$, ($b$) coherent signal generated by a pair of molecules $\alpha\neq\beta$. Eqs. (\ref{eq:Wcoh1}) - (\ref{eq:Pol}) (coherent) and Eqs. (\ref{eq:Winc1}) - (\ref{eq:Ppr}) (incoherent) can be read off these diagrams (for diagram rules see Ref. \cite{Rah10}). The shaded area represents an excitation by an arbitrary sequence of pulses, which prepares the molecule in a superposition state.}
\label{fig:det}
\end{figure}

Time-and-frequency gated spontaneous light emission signals is given by an integrated intensity of the electric field of an emitted photon. It may be further represented by the spectral and temporal overlap of a bare signal $W_B$ and a detector spectrogram $W_D$ \cite{Dor12}
\begin{equation}\label{eq:S011}
S(\bar{t},\bar{\omega})=\int_{-\infty}^{\infty}dt'\frac{d\omega'}{2\pi}W_D(\bar{t},\bar{\omega};t',\omega')W_B(t',\omega').
\end{equation}
The detector offers a window of observation centered at time $\bar{t}$ and frequency $\bar{\omega}$. Even though $\bar{t}$ and $\bar{\omega}$ can be varied independently the actual temporal $\sigma_T$ and spectral $\sigma_\omega$ resolutions are not independent and satisfy $\sigma_T\sigma_\omega\geq1$. This is guaranteed by Eq. (\ref{eq:S011}). We shall assume a point-size detector, therefore omitting all effects of spatial resolution only retaining temporal and spectral gating \cite{Dor12}. Eq. (\ref{eq:S011}) was introduced originally as signal processing for the classical field  \cite{Sto94} but it more broadly applies to quantum fields \cite{Dor12}. The bare signal $W_B$ contains all of the relevant information about the externally driven molecules and is given by a correlation function of matter that further includes a sum over the detected modes. In Appendix \ref{app:mic} we present superoperator expressions for $W_B$ of a system prepared in an arbitrary non stationary state without treating the preparation explicitly. The bare signal (\ref{eq:WE02}) represented by the closed path time-loop diagram shown in Fig. \ref{fig:det}a,b is given by time-ordered Green's functions for superoperators  (TIGERS)  \cite{Har08}. It contains a double summation over molecular positions $\mathbf{r}_\alpha$ and $\mathbf{r}_\beta$ and contains two types of contributions: incoherent emission terms with $\alpha=\beta$ (Fig. \ref{fig:det}a) and coherent emission $\alpha \ne \beta$ (Fig. \ref{fig:det}b). 

We start with the coherent emission which is induced by pairs of molecules labeled $\alpha$ and $\beta$. Here we can factorize the matter correlation in (\ref{eq:WE02})  $\langle \mathcal{T}\hat{V}^{\alpha}_L(t_1')\hat{V}_R^{\beta\dagger}(t_1)\rangle \to\langle \hat{V}^{\alpha}_L(t_1')\rangle \langle\hat{V}_R^{\beta\dagger}(t_1)\rangle$. $W_B$ is then factorized into a product of amplitudes (see Appendices \ref{app:super}, \ref{app:Lar})
\begin{align}\label{eq:Wcoh1}
&W_{Bcoh}(t',\omega')=A^2\sum_{\mu}\int_0^{\infty}d\tau e^{-i\omega'\tau}\sum_{\alpha=1}^N\sum_{\beta=1}^{N-1}\frac{e^{i\mathbf{k}_n\cdot(\mathbf{R}_\beta-\mathbf{R}_\alpha)}}{R_\alpha R_\beta}\notag\\
&\times \ddot{P}^{(\alpha,\mu)}(t'-\tau/2-R_\alpha/c)\ddot{P}^{(\beta,\mu)*}(t'+\tau/2-R_\beta/c).
\end{align}
Here the polarization of molecule $\alpha$ in the cartesian direction $\mu$  is given by
\begin{align}\label{eq:Pol}
&P^{(\alpha,\mu)}(t)=\notag\\
&\sum_{a,b}\rho_{aa}\langle\langle bb| \mathcal{T}V_{\mu L}(t)\exp\left(-\frac{i}{\hbar}\int_{-\infty}^{t}H'_-(T)dT\right)|aa\rangle\rangle_\alpha,
\end{align}
where $\rho_{aa}$ represents the initial equilibrium density matrix, $\mathcal{T}$ is responsible fo time ordering of superoperators. When using a purely temporal gate, $F_f(\omega,\bar{\omega})=1$  and we get $W_D(\bar{\omega},\bar{t};t,\tau)=\delta(\tau)F_t^{*}(t+\tau/2,\bar{t})F_t(t-\tau/2,\bar{t})$, where $W_D(\bar{\omega},\bar{t};t,\tau)=\int_{-\infty}^{\infty}\frac{d\omega}{2\pi}e^{i\omega\tau}W_D(\bar{\omega},\bar{t};t,\omega)$. For an ideal time gate $|F_t(t,\bar{t})|^2=\delta(t-\bar{t})$  and Eqs. (\ref{eq:S011}) and (\ref{eq:Wcoh1}) result in the time-resolved signal
\begin{equation}\label{eq:St}
S(\bar{t})=|E(\bar{t})|^2.
\end{equation}
i.e.
\begin{align}\label{eq:Wct}
&S_{coh}(\bar{t})=A^2\sum_{\mu}\Big|\sum_{\alpha=1}^N\frac{e^{-i\mathbf{k}_n\cdot\mathbf{R}_\alpha}}{R_\alpha}\ddot{P}^{(\alpha,\mu)}(\bar{t}-R_\alpha/c)\Big|^2,
\end{align}
where $A$ is a normalization constant and the double-dot represents the second time derivative. In the opposite, pure frequency, gating, i.e.  $F_t(t,\bar{t})=1$, and the frequency gate is very narrow, such that $F_f(t,\bar{\omega})=\frac{\sqrt{\gamma}}{\pi} e^{-i\bar{\omega}t-\gamma t}\theta(t)$ at $\gamma\to0$, then $W_D(\bar{\omega},\bar{t};\omega,\tau)=e^{-i\bar{\omega}\tau}$. We then obtain the frequency-resolved signal
\begin{equation}\label{eq:Sw}
S(\bar{\omega})=|E(\bar{\omega})|^2
\end{equation} 
where $E(\omega)=\int_{-\infty}^{\infty}dtE(t)e^{i\omega t}$. Using Eq. (\ref{eq:Wcoh1}) we obtain
\begin{align}\label{eq:Wcw}
&S_{coh}(\bar{\omega})=A^2\sum_{\mu}\Big|\sum_{\alpha=1}^N\frac{e^{i(\bar{\omega} R_\alpha/c-\mathbf{k}_n\cdot\mathbf{R}_\alpha)}}{R_\alpha}\bar{\omega}^2 P^{(\alpha,\mu)}(\bar{\omega})\Big|^2,
\end{align}
It follows from the Larmor formula Eqs. (\ref{eq:Wct})  or (\ref{eq:Wcw}) that only when the measurement is either solely time or frequency-gated, can the coherent signal be expressed in terms of the modulus square of a matter amplitude. The more general time and frequency gated signals (\ref{eq:WE01}), (\ref{eq:WE02}) and (\ref{eq:Wcoh1}) do not have this form. 

We next turn to the incoherent signal (Fig. \ref{fig:det}a) originating from processes in which both signal-mode interactions occur with the same molecule $\alpha=\beta$. This signal is not given by a modulus square of the polarization amplitude. However by expanding it in the molecular eigenstates we can express it in terms of other type of amplitudes. After a bit of algebra (see Appendix \ref{app:Lar} for details) we obtain
\begin{align}\label{eq:Winc1}
&W_{Binc}(t',\omega')=A^2\sum_{\mu}\int_0^{\infty}d\tau e^{-i\omega'\tau}\sum_{\alpha=1}^N\frac{1}{R_\alpha^2}\sum_{a,b}\rho_{aa}\notag\\
&\times\ddot{T}_{ba}^{(\alpha,\mu)}(t'-\tau/2-R_\alpha/c)\ddot{T}_{ab}^{(\alpha,\mu)*}(t'+\tau/2-R_\alpha/c).
\end{align}
Here 
\begin{align}\label{eq:Ppr}
T_{ba}^{(\alpha,\mu)}(t)=\langle b(t)|\mathcal{T}V_\mu(t)\exp\left(-\frac{i}{\hbar}\int_{-\infty}^{t}H'(T)dT\right)|a\rangle_\alpha
\end{align}
represents a matter transition amplitude of molecule $\alpha$ starting with initial state $a$ and reaching the final state $b$ at time $t$.

For a pure time domain gating (\ref{eq:St}), Eq. (\ref{eq:Winc1}) gives
\begin{align}\label{eq:Winct}
&S_{inc}(\bar{t})=A^2\sum_{\mu,a,b}\sum_{\alpha=1}^N\frac{1}{R_\alpha^2}|\ddot{T}_{ba}^{(\alpha,\mu)}(\bar{t}-R_\alpha/c)|^2,
\end{align}
Pure frequency gating (\ref{eq:Sw}) yields
\begin{align}\label{eq:Wincw}
&S_{inc}(\bar{\omega})=A^2\sum_{\mu,a,b}\sum_{\alpha=1}^N\frac{1}{R_\alpha^2}\bar{\omega}^4|T_{ba}^{(\alpha,\mu)}(\bar{\omega})|^2.
\end{align}

Eqs. (\ref{eq:Winc1}) - (\ref{eq:Wincw}) indicate that the incoherent signal may not be generally recast in the form of square of the total polarization alone Eq. (\ref{eq:Pol}). Rather, it depends on a different set of  matter quantities (\ref{eq:Ppr}), transition amplitudes, that represent the possible quantum pathways of matter from state $a$ to state $b$. This calculation does not require superoperators since the amplitudes represent the evolution of the bra or ket separately and can be recast in Hilbert space. The polarization (\ref{eq:Pol}) on the other hand is governed by evolution of both bra and ket and therefore can only be defined using superoperators.

The absence of spatial phase factors in Eq. (\ref{eq:Winc1}) indicates that the incoherent emission is isotropic and is independent on the wave vector of the incoming pulses. The single summation over molecules yields linear $\sim N$-scaling with the number of active molecules. The coherent signal  (\ref{eq:Wcoh1}) in contrast scales as $N(N-1)$. This signal is directional and propagates according to the phase matching condition $\mathbf{k}_s=\mathbf{k}_{n}$. Clearly, the coherent and incoherent signals can be always distinguished by their $N(N-1)$ vs $N$-scaling and  directionality vs isotropic emission. A different type of size scaling exists in hyper-Rayleigh techniques. Consider a sample of $N$ particles each composed of $M$ molecules. Assuming coherent intraparticle and incoherent interparticle scattering the signal will then scale as $NM(M-1)$ and will appear to be linearly scaling in the number of particles. $N$-scaling can be similarly obtained if each molecule $\alpha$ is near some set of $N'$ neighbor molecules $\beta$ such that $|\mathbf{k}_s|\cdot|\mathbf{r}_{\alpha}-\mathbf{r}_\beta|\ll1$, then it will emit in phase with this set of molecules, thus lending an overall factor of $NN'$. Incoherent sum frequency generation is an example of hyper-Rayleigh signal \cite{Ter65,Cla91,Cla93}. This signal scales linearly with the number of molecules similar to the incoherent fluorescence.

Classically, spontaneous light emission from a non-stationary state is calculated by the Larmor formula (see Appendix \ref{app:sc}). All the necessary information is then contained in the charge density of the accelerating particle and the radiation power is related to the second order time derivative of the macroscopic polarization. The classical derivation apparently applied even to a single molecule and the result is then simply multiplied by the number of molecules. This yields linear $N$-scaling. It follows from our analysis that the classical formula is valid only for the coherent signal from a mesoscopic volume and may not be applied to a single molecule. Light scattering from a single molecule is incoherent and provides new information about quantum pathways of matter. It may not be recast using the charge density alone.

To further evaluate the sum over molecules in Eqs. (\ref{eq:Wcoh1}) - (\ref{eq:Wincw}) we first consider small samples (compared to the optical wavelength) when retardation effects can be neglected. In this case both coherent and incoherent signals are independent of the direction of the incoming pulses. Straightforward analysis yields for the incoherent signal $\sum_\alpha R_\alpha^{-2}\sim N/R_c^2$, where $R_c$ is characteristic size of the sample. Similarly since $|\mathbf{k}_n|\cdot|\mathbf{r}_{\alpha}-\mathbf{r}_\beta|\ll1$ the coherent component yields $\sum_\alpha\sum_\beta (R_{\alpha}R_\beta)^{-1}\sim NN'/R_c^2$. Thus, both signals scale linearly with molecular density and inverse proportional to the square of the sample size. For extended samples, when retardation becomes important, the coherent signal yields a directional emission with $N^2$-scaling: $\sum_{\alpha,\beta}e^{i(\omega R_\alpha/c-\mathbf{k}_n\cdot\mathbf{R}_\alpha)-i(\omega'R_\beta/c-\mathbf{k}_n\cdot\mathbf{R}_\beta)}(R_\alpha R_\beta)^{-1}\simeq N(N-1)\delta(\omega/c-k_n)\delta(\omega'/c-k_n)$.

When retardation is neglected, the coherent signal  (\ref{eq:Wct}) can be recast in the Larmor form. The most striking limitation  is that this form may not be applied to the single molecule case, as shown in Eq. (\ref{eq:Winct}); it requires  pairs of molecules. The conventional classical derivations define the signal for a single molecule and then multiply it by the number of molecules. We showed that the $N$-scaling can be realized in the coarse grained system with mesoscopic grains via hyper-Rayleigh scattering where the short-range coherence makes the signal to scale $N$ and not $N^2$.

\section{Light Scattering Off a Non-Stationary State}

We now calculate the incoherent and coherent signals in a multilevel system prepared in an arbitrary superposition state. In Section II, the evolution governed by $H_-'(t)$ and $H'(t)$ in Eqs. (\ref{eq:Pol}) and (\ref{eq:Ppr}), respectively caused by an arbitrary sequence of pulses which occurs prior to the last emission constitutes the preparation process that leaves the system in the superposition state. Here we ignore the details of the preparation (the shaded area in Fig. \ref{fig:det}) and start our analysis after that time period where the system has been prepared in non stationary superposition state: $\rho_{cc'}$ in Fig. \ref{fig:det}a and two molecules in $\rho_{cb}$ and $\rho_{bc'}$, respectively, in Fig. \ref{fig:det}b. 

\subsection{The coherent signal}

Starting with Eq. (\ref{eq:Pol}) and omitting  the superscript $\alpha$. For clarity we can recast the $\nu$-th  component of the polarization in the form
\begin{equation}
P^{(\nu)}(t)=Tr[V_\nu\rho(t)].
\end{equation}
After multiple interactions with incoming pulses the matter density matrix Fig. \ref{fig:det}b is $\rho_{cb}$. We further define $\mu_{cb}^{(\nu)}\equiv\langle b\vert V_\nu\vert c\rangle$. We can then write:
\begin{equation}
P^{(\nu)}(t)=\sum_{b,c}\rho_{bc}\mu_{cb}^{(\nu)}e^{-i\omega_{bc}t}.
\end{equation}
Assuming that molecules $\alpha$, $\beta$ are identical and neglecting retardation we obtain the following compact form for the bare spectrogram:
\begin{align}\label{eq:wcoh3}
&W_{B,coh}(t',\omega')\notag\\
&=N(N-1)\tilde{A}^2\int_0^{\infty}d\tau e^{-i\omega'\tau}\ddot{P}(t'-\tau/2)\ddot{P}^{*}(t'+\tau/2),
\end{align}
where we have further averaged over the random dipole orientation  $\sum_\nu P^{(\nu)}(t_1)P^{(\nu)*}(t_2)\simeq\frac{1}{3}P(t_1)P^{*}(t_2)$ yielding $\tilde{A}=A/\sqrt{3}$. Since $P(t)$  only depends  on a single interaction with the signal mode, the coherent signal governed by the expectation value of the signal field is finite and the molecule returns to state $|b\rangle\langle b|$ at the end of the process. In the case of ideal time-gating, we recover the Larmor formula:
\begin{align}\label{eq:scoht}
S_{coh}(\bar{t})=N(N-1)\frac{\tilde{A}^2}{2}\lvert\ddot{P}(\bar{t})\rvert^2,
\end{align}
where
\begin{align}
\ddot{P}(t)=\sum_{b,c}\rho_{bc}\omega_{bc}^2\mu_{cb}^{(\nu)}e^{-i\omega_{bc}\bar{t}}.
\end{align}
For purely frequency gating, we similarly get:
\begin{align}\label{eq:scohw}
S_{coh}(\bar{\omega})=N\tilde{A}^2\sum_{b,c}\omega^4_{bc}\lvert\mu_{cb}\rvert^2\lvert\rho_{bc}\rvert^2\delta(\bar{\omega}-\omega_{bc}).
\end{align}

\begin{figure}[t]
\begin{center}
\includegraphics[trim=0cm 0cm 0cm 0cm,angle=0, width=0.4\textwidth]{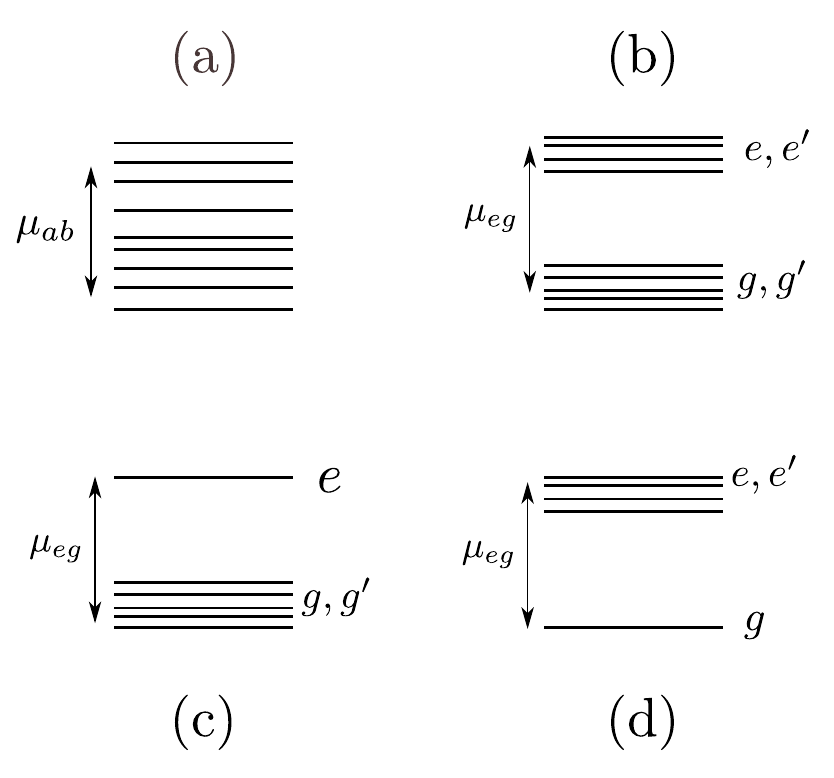}
\end{center}
\caption{Various valence level-schemes considered in this paper. ($a$) - a general level scheme with arbitrary transition dipoles. ($b$) - a two band model with only interband transition dipole,  (c) - two bands with a single excited state. ($d$) - two bands with a single ground state.}
\label{fig:levels}
\end{figure}

\subsection{The incoherent signal}

To calculate the $\alpha=\beta$ terms in  Eq.(\ref{eq:WE02}) (see Fig. \ref{fig:det}a) we must evaluate the following matter correlation function:

\begin{align}
\langle\hat{V}_L(t_1')\hat{V}_R^{\dagger}(t_1)\rangle=Tr\left[\hat{V}^{\dagger}(t_1)\rho\hat{V}(t_1')\right] \\ \notag
=\sum_{b,c,c'}\rho_{c,c'}\langle b|\hat{V}(t_1')|c\rangle\langle c'|\hat{V}^{\dagger}(t_1)|b\rangle\\ \notag=\sum_{b,c,c'}\rho_{c'c}\mu_{cb}\mu^*_{c'b}e^{-i\omega_{bc}t'_1}e^{i\omega_{bc'}t_1},
\end{align}
where the state of the system prior to the emission is given by the nonstationary density matrix element $\rho_{c'c}$ and the emission brings it to the population state $|b\rangle\langle b|$. Combining this with Eqs. (\ref{eq:Winc1}) gives:
\begin{align}\label{eq:Wbincss}
W_{B,inc}(t',\omega')=\pi \tilde{A}^2 N \sum_{b,c,c'}\omega_{bc}^2\omega_{bc'}^2\mu_{cb}\mu^*_{c'b}\rho_{c'c}\notag \\ 
\times e^{i(\omega_{bc'}-\omega_{bc})t'}\delta(\omega'-\frac{\omega_{bc}+\omega_{bc'}}{2}).
\end{align}
The incoherent bare spectrogram (\ref{eq:Wbincss}) may not be recast in the Larmor form (Eq. (\ref{eq:wcoh3})). The time gated incoherent signal is
\begin{align}\label{eq:sinct1}
S_{inc}(\bar{t})=\frac{N\tilde{A}^2}{2}\sum_{b,c,c'}\omega_{bc}^2\omega_{bc'}^2 \mu_{cb}\mu^*_{c'b}\rho_{c'c}e^{i(\omega_{bc'}-\omega_{bc})\bar{t}}.
\end{align}
The frequency-gated signal is:
\begin{align}\label{eq:sincw1}
S_{inc}(\bar{\omega})=N\tilde{A}^2\sum_{b,c}\omega_{bc}^4\mu^2_{cb}\rho_{cc}\delta(\bar{\omega}-\omega_{bc})
\end{align}
In contrast with the coherent signal, we see that an excited-state population rather than a coherence in the relevant transitions is required to produce a signal.  This is because all interactions are with a single molecule and the radiation signal mode (initially in a vacuum state) must be brought to a population one photon state to generate a signal.

\begin{figure}[t]
\begin{center}
\includegraphics[trim=0cm 0cm 0cm 0cm,angle=0, width=0.48\textwidth]{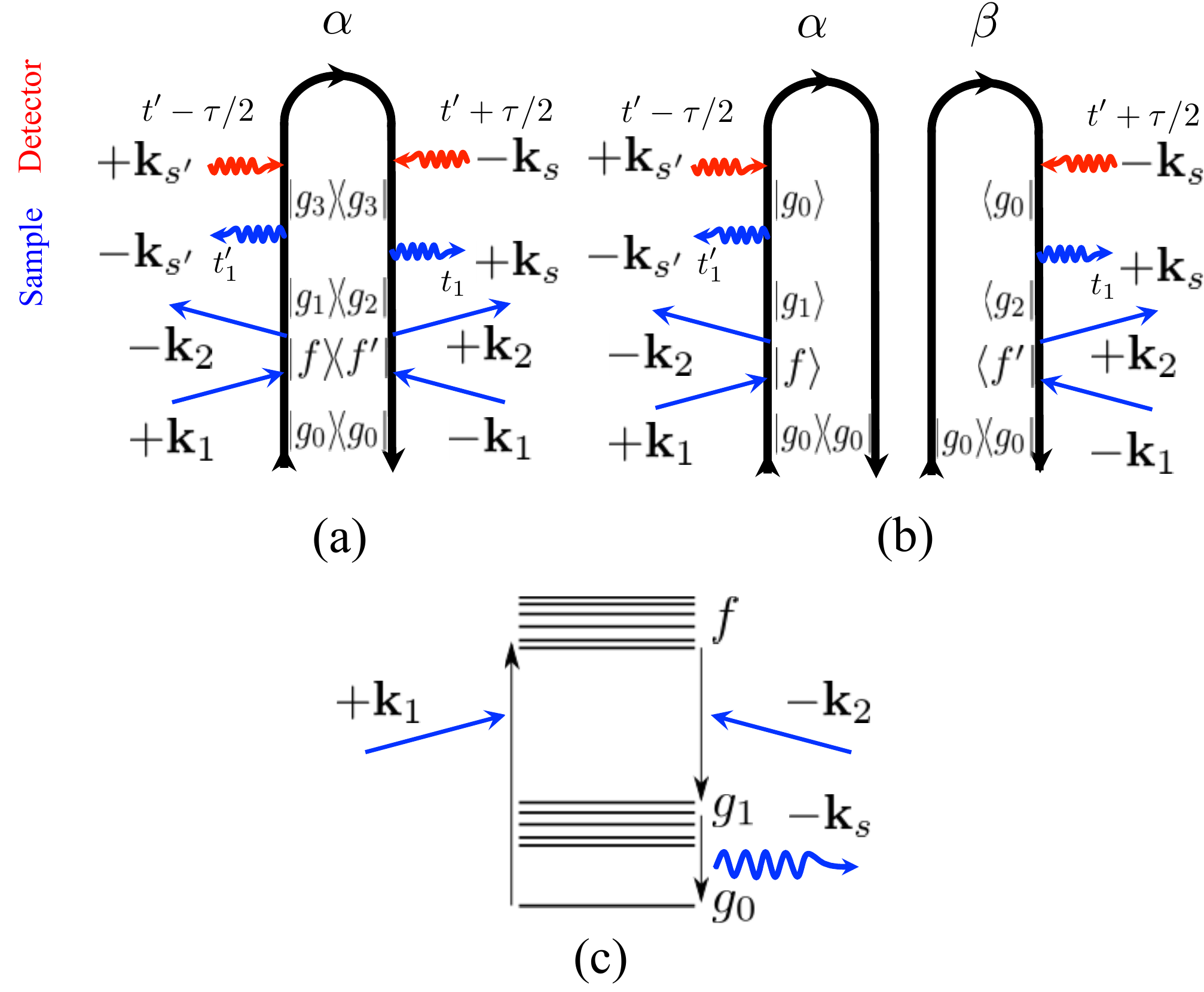}
\end{center}
\caption{(Color online) The stimulated X-ray Raman induced NLS process. Straight arrows correspond to interactions with X-ray pulses, wavy arrows represent spontaneous emission. ($a$) loop diagram for the bare SRIF signal ($b$) the bare coherent DFG  signal, (c) The level scheme used for cysteine is composed of a ground state $g_0$ and 50 valence excited states $g_1$ with energies between $\sim$5.75eV and $\sim$11.5eV, $f$ are core excited states.}
\label{fig:detR}
\end{figure}

We now apply our results a level scheme composed of two manifolds of states with interband dipole elements as shown in Fig. \ref{fig:levels}b. Eqs. (\ref{eq:wcoh3}) - (\ref{eq:Wbincss}) which gives:
\begin{widetext}
\begin{align}\label{eq:wbare3}
W_{B0,coh}(t',\omega')&=\pi N\tilde{A}^2\sum_{ee',gg'}\omega_{eg}^2\omega_{e'g}^2\mu_{ge}^{*}\mu_{g'e'}\rho^*_{ge}\rho_{g'e'}e^{i(\omega_{eg}-\omega_{e'g'})t'}\delta\left(\omega'-\frac{\omega_{eg}+\omega_{e'g'}}{2}\right)
\end{align}
\begin{align}\label{eq:wbare4}
W_{B0,inc}(t',\omega')&=\pi N\tilde{A}^2\sum_{gee'}\omega_{eg}^2\omega_{e'g}^2\mu_{ge}^{*}\mu_{ge'}\rho_{ee'}e^{i(\omega_{eg}-\omega_{e'g})t'}\delta\left(\omega'-\frac{\omega_{eg}+\omega_{e'g}}{2}\right),
\end{align}
\end{widetext}
where the subscript ``0" signifies the two-band model.  In the case of time-and-frequency gating, various transitions between ground and singly-excited states become coupled.  A time-gated measurement (\ref{eq:St}) yields
 \begin{align}\label{eq:S3ct}
S_{0,coh}(\bar{t})&=N\frac{\tilde{A}^2}{2}\left[\sum_{eg}\omega_{eg}^2\mu_{ge}\rho_{ge}e^{-i\omega_{eg}\bar{t}}\right]^2
\end{align}
\begin{align}\label{eq:S3qt}
S_{0,inc}(\bar{t})&=N\frac{\tilde{A}^2}{2}\sum_g\sum_{e,e'}\omega_{eg}^2\omega_{e'g}^2\mu_{ge}\rho_{e'e}e^{i(\omega_{e'g}-\omega_{eg})\bar{t}}.
\end{align}
For an ideal frequency gating (\ref{eq:Sw}), the signal (\ref{eq:wbare3}) - (\ref{eq:wbare4}) reads
 \begin{align}\label{eq:S3cw}
 S_{0,coh}(\bar{\omega})&=N\tilde{A}^2\sum_{eg}\omega_{eg}^4|\mu_{ge}|^2|\rho_{ge}|^2\delta(\bar{\omega}-\omega_{eg})
 \end{align}
 \begin{align}\label{eq:S3qw}
 S_{0,inc}(\bar{\omega})&=N\tilde{A}^2\sum_{eg}\omega_{eg}^4|\mu_{ge}|^2\rho_{ee}\delta(\bar{\omega}-\omega_{eg})
 \end{align}

Eqs. (\ref{eq:S3cw}) and (\ref{eq:S3qw}) are very similar, both given in the form of the Fermi golden rule for spectrally well separated states. However, this is no longer the case for time-domain measurements. In Eq. (\ref{eq:S3ct}), the initial and final states enter in the same way (the summations are fully interchangeable and all transitions add at the amplitude level) while in Eq. (\ref{eq:S3qt}), a trace is taken over final states after adding the amplitudes for the transitions to a given final state.  The consequences of this difference are most readily seen by comparison of the two level schemes shown in Fig. \ref{fig:levels}.  In case $(c)$, there is a single excited state and a manifold of ground states.  In case $(d)$, the situation is reversed and there is a manifold of excited states and a single ground state. Both cases will generate a beating term in the coherent signal.  In contrast, the incoherent signal (\ref{eq:S3qt}) gives a beating term for case $(d)$ but not for case $(c)$.  $V$ and $\Lambda$ type three-level model of atoms  are commonly discussed in quantum optics \cite{scully, dor11}; these correspond to our cases $(d)$ and $(c)$ respectively. The total signal is $S=S_{coh}+S_{inc}$. $S_{coh}$ only has beats for ($d$). The semiclassical approach only gives $S_{coh}$.


\section{Difference Frequency Generation VS Fluorescence Induced by a Stimulated X-ray Raman Process}

We have simulated these signals for cysteine  -  an amino acid which serves as an important structural unit in connecting different regions of proteins by disulfide bonds. Following excitation by stimulated X-ray Raman process resonantly tuned to either the nitrogen, oxygen or the sulfur K-edges. The signal is given by the diagrams of Fig. \ref{fig:detR} and involves six radiation/matter interactions. In this case the incoherent signal is SRIF the coherent signal is DFG as depicted in Fig. \ref{fig:detR}c. An attosecond X-ray pulse excites the core transition to state $f$ by absorbing a $\mathbf{k}_1$ photon which is then de-excited with $-\mathbf{k}_2$ photon, leaving the system in a superposition of valence states $g_1$ and $g_2$ with amplitudes given by Eq. (A7) of Ref. \cite{biggs}. This results in a weakly-excited, pure-state $g_3$.  We thus expect the incoherent and coherent signals to coincide for the $g_{1,2}\leftrightarrow g_0$ transition. The relevant diagrams are shown in Fig. \ref{fig:detR}a,b. For a two state model system prepared in a pure-state the state vector reads
\begin{equation}\label{eq:psi2}
|\psi_{0}\rangle=\kappa_g|g\rangle+\sum_e\kappa_e|e\rangle,
\end{equation} 
where expressions for $\kappa_j$, $j=e,g$ in Eq. (\ref{eq:psi2}) are given in Ref. \cite{biggs}. In the limit of weak excitation most of the population is in the ground state $|\kappa_g|^2\sim1$ and $|\kappa_e|^2\ll 1\forall e$ (the excitation is perturbative).  Thus, Eq. (\ref{eq:scohw}) reads:
\begin{equation}
S_{coh}(\bar{\omega})=N\tilde{A}^2\sum_{b,c}\omega_{bc}^4|\mu_{cb}|^2|\kappa_b|^2|\kappa_c|^2\delta(\bar{\omega}-\omega_{bc})
\end{equation}
Since $|\kappa_e|^2\ll 1$ and $|\kappa_g|^2\sim1$ the leading contribution yields
\begin{equation}
S_{coh}(\bar{\omega})=N\tilde{A}^2\sum_{c}\omega_{gc}^4|\mu_{cg}|^2|\kappa_c|^2\delta(\bar{\omega}-\omega_{gc}),
\end{equation}
where we assumed $b=g$. This is clearly a subset of the incoherent signal (Eq. (\ref{eq:sincw1})) which, under the assumption of a pure state, is given by:
\begin{equation}
S_{inc}(\bar{\omega})=N\tilde{A}^2\sum_{b,c}\omega_{bc}^4|\mu_{cb}|^2|\kappa_b|^2\delta(\bar{\omega}-\omega_{bc})
\end{equation}
Thus, for a weakly excited system in a pure state, the incoherent and coherent signals coincide for transitions from the excited states to the ground state.  When the excited-state manifold bandwidth is smaller than the band gap, these transitions appear in the high-energy part of the emission spectra (see Section V).  The intraband transitions within the excited state manifold  then appear in the red part of the spectra and only show up in the incoherent signal. 

\par

\begin{figure*}
\includegraphics[width=0.8\textwidth]{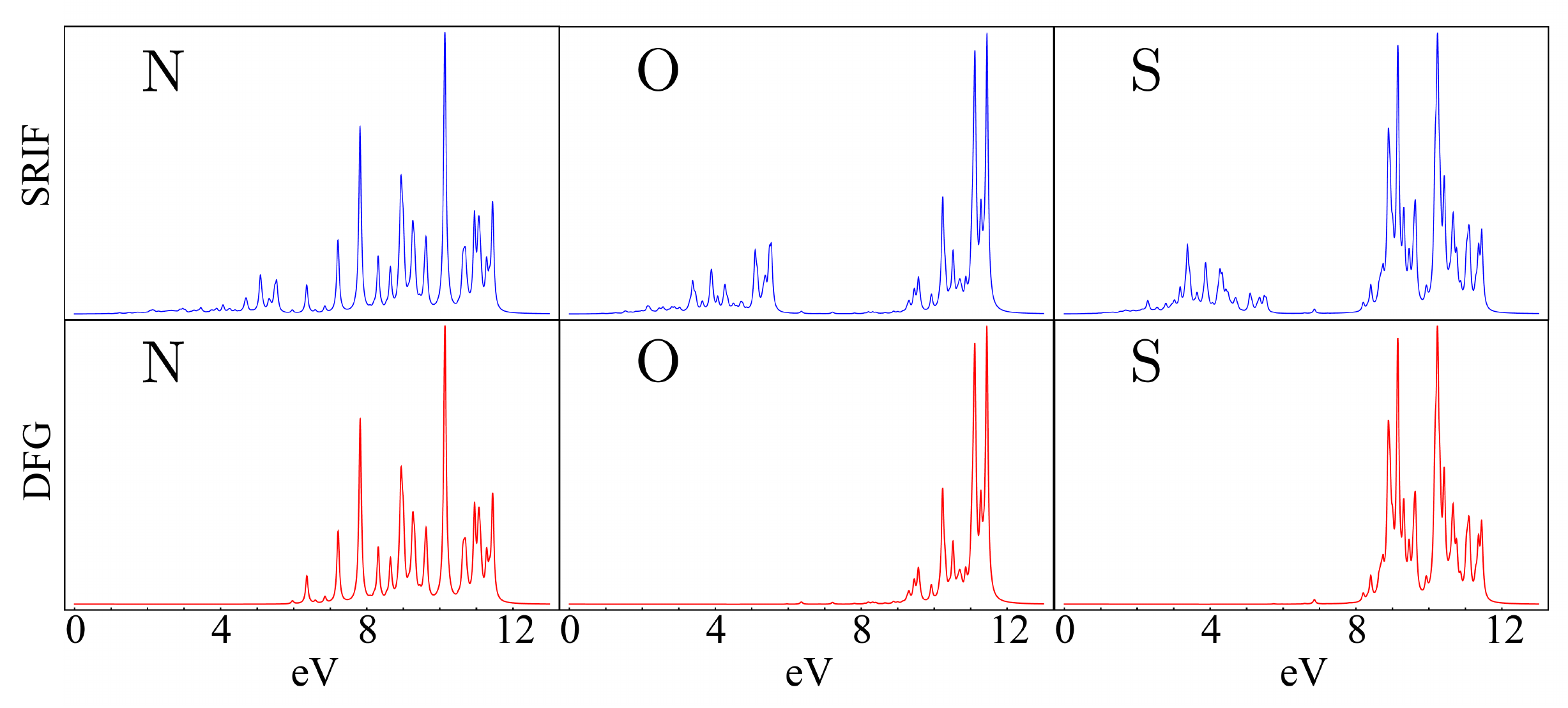}
\caption{(Color online) SRIF (Top) and DFG (Bottom) NLS signals of cysteine following stimulated Raman excitation by an X-ray pulse of width 14.2 eV and central frequency tuned to the nitrogen ($\sim 404.4$ eV), oxygen ($\sim 532.2$ eV), or sulfur ($\sim 2473.5$ eV) K-edges, as indicated. The two signals coincide for $g_{1,2}\leftrightarrow g_0$ transitions. Transitions between excited states  $g_{1,2}\leftrightarrow g_3$, $\bar{\omega}<5$ eV do not show up in the coherent signal. $\Gamma\sim$0.04 eV is used for the gating bandwidth.}\label{fig:cohinc}
\end{figure*}

 The optimized geometry was obtained with the quantum chemistry package Gaussian09 \cite{G09} at the B3LYP\cite{Becke93,SDCF94}/6-311G** level. Core excited states were calculated with restricted excitation window time-dependent density functional theory (REW-TDDFT)\cite{Niri} implemented in NWChem code \cite{NWChem}. Core and valence energy levels and transition dipoles moments were calculated with a locally modified version of NWChem code at the CAM-B3LYP\cite{YTH04}/6-311G** level of theory and within the Tamm-Dancoff approximation\cite{HHG99b}. Additional computational details are given in Ref. \cite{ZBHG12}. The calculated  frequency gated SRIF (\ref{eq:sincw1}) and DFG (\ref{eq:scohw}) signals obtained by excitation of the nitrogen, oxygen and sulfur K edges are shown in the top and bottom rows respectively of Fig. \ref{fig:cohinc}.

\begin{figure}[t]
\includegraphics[width=.35\textwidth]{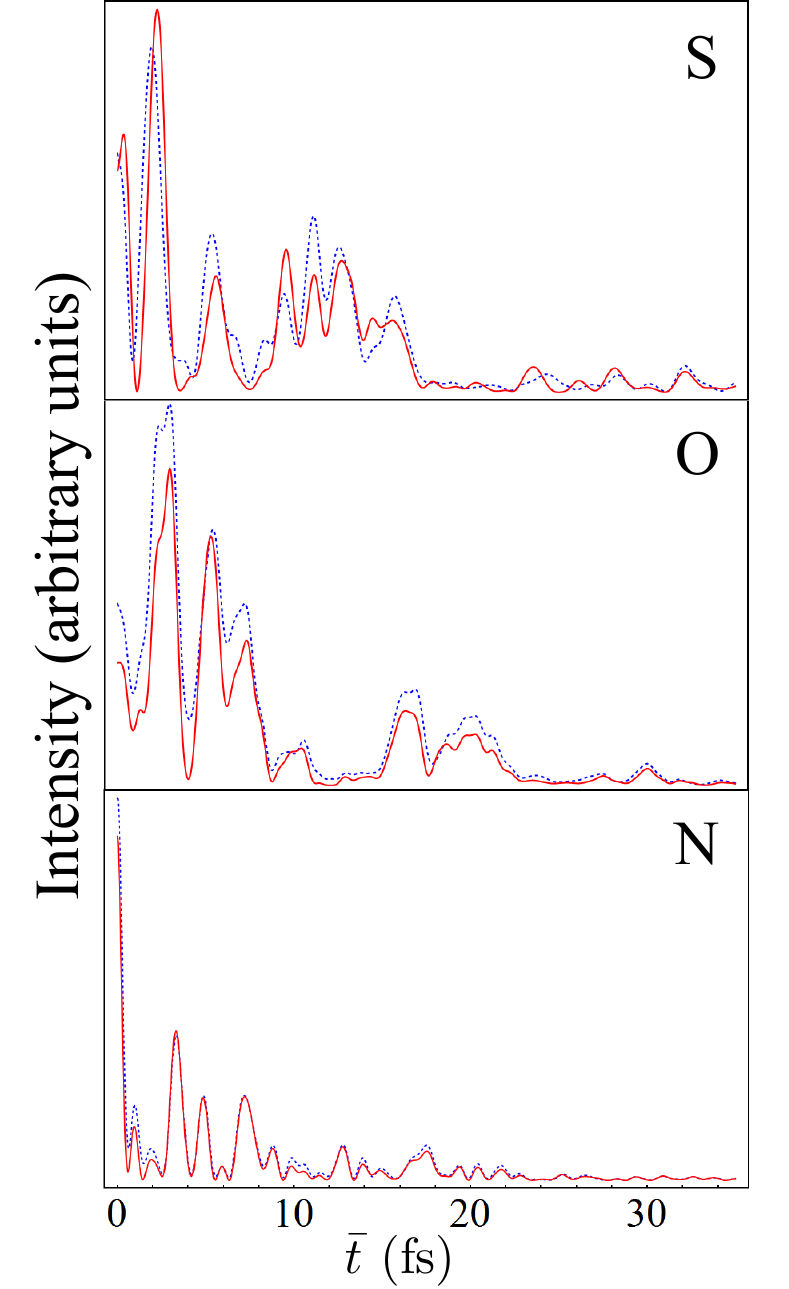}
\caption{(Color online) Time resolved SRIF - solid red line  and DFG - dotted blue line signals are compared for X-ray pulses resonant with $S$, $O$ and $N$ in cysteine. $\Gamma\sim$0.04 eV is used for the gating bandwidth.}\label{fig:timedom}
\end{figure}

\begin{figure}
\includegraphics[width=0.45\textwidth]{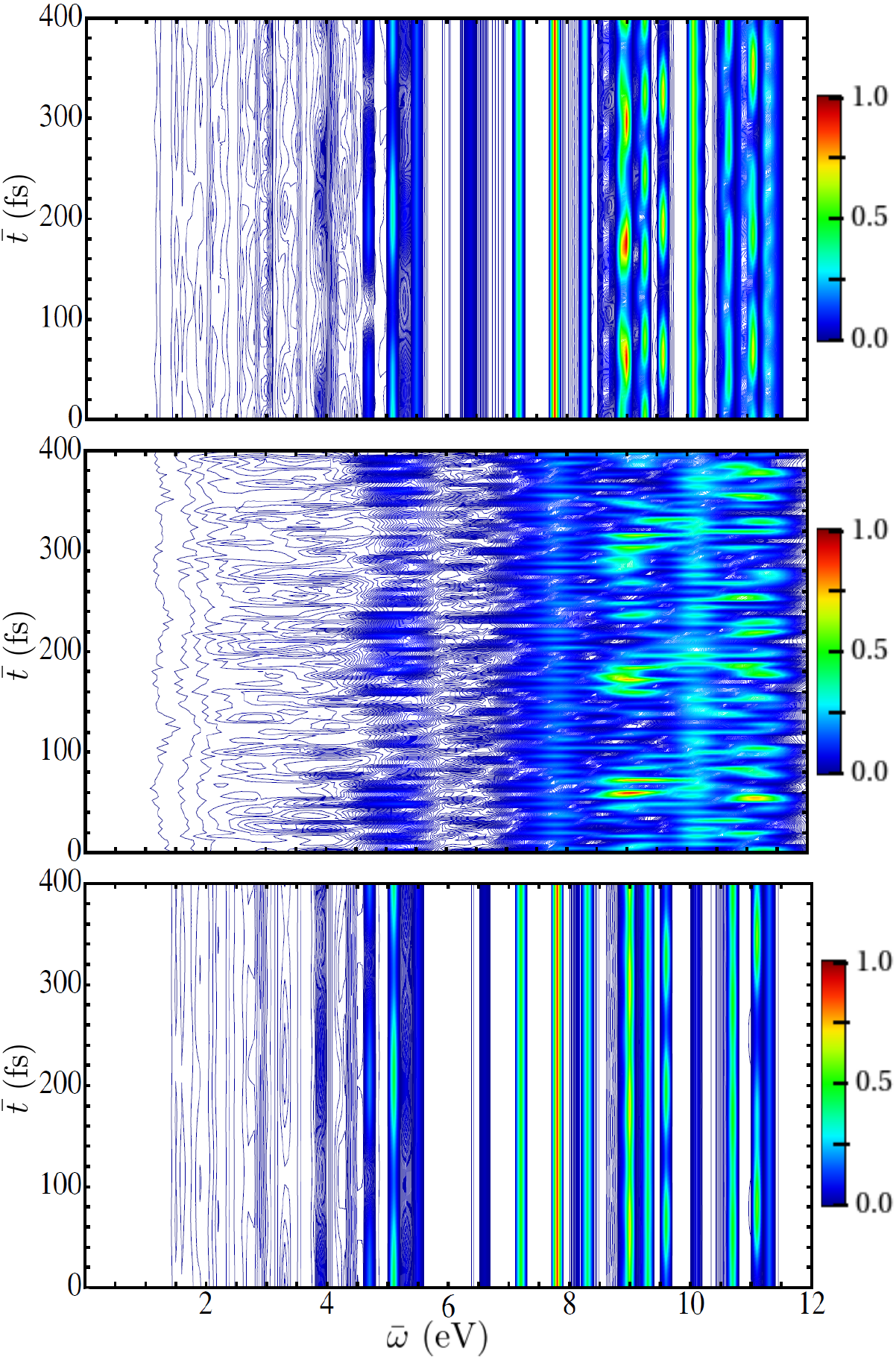}
\caption{(Color online) Time and frequency gated SRIF signals of cysteine (Eq. (\ref{eq:S011})). The gating parameters in atomic units $\sigma_T=1000$, $\sigma_{\omega}=0.001$ - ($a$), $\sigma_T=100$, $\sigma_{\omega}=0.01$ - ($b$), and $\sigma_T=2000$, $\sigma_{\omega}=0.005$ - ($c$). These permit all transitions within $\sim 0.054$ eV, $0.54$eV and $0.1$ eV of each other to interfere, respectively. Top panel has optimal gating parameters and reveals both areas of low transition density (where the intensities are time independent) and areas of higher densities (where beats develop as a result of interference between transitions).  Note the particularly prominent beating near $9$ eV with an approximate period of $120$ fs.  The actual distance between these states is $\sim 0.0126$ Ha well within the allowed detection bandwidth. Middle panel  has low frequency resolution but rather high temporal resolution that results in clear beating signal. The bottom panel has high spectral  but  low temporal resolution, that results in the suppression of formerly prominent beating at $\sim 9$ eV.  Further narrowing the frequency-domain gating widths eliminates it altogether.}
\label{fig:2D}
\end{figure}

We first note that the SRIF and DFG signals coincide in the high-frequency regime.  The lowest excited state has an energy $\sim 5.74$ eV while the excited state manifold bandwidth is $\sim 5.70$ eV.  These high-energy transitions are from the various excited states to the ground state. The low-energy $\omega< 5$ eV features that only appear in the incoherent spectra represent transitions between excited states $g_{1,2}\leftrightarrow g_3$. The frequency gated coherent and incoherent signals are very different. However the corresponding time gated signals shown in Fig. \ref{fig:timedom} are less distinct. 

Fig. \ref{fig:2D} depicts the time-and-frequency gated spectrograms of the SRIF signal computed by Eq. (14) of Ref. \cite{Dor12} with gaussian gating functions
\begin{align}
F_f(\omega,\bar{\omega})=e^{-\frac{(\omega-\bar{\omega})^2}{2\sigma_\omega^2}},\quad F_t(t,\bar{t})=e^{-\frac{(t-\bar{t})^2}{2\sigma_T^2}},
\end{align}
where  $\sigma_\omega$ and $\sigma_T$ are the corresponding bandwidths of the frequency and time gates, respectively. These obey Fourier uncertainty relation $\sigma_\omega\sigma_T\geq1$.  Some transitions in Fig. \ref{fig:2D}a do not vary with time while others beat.  Beating occurs because of interference with nearby transitions and thus indicates a higher density of transitions.  The beating period gives the interval between the interfering states $a,b$ via $\vert\omega_{ab}\vert=\frac{2\pi}{T_{beat}}$.  The gating parameters determine the maximum energy difference (and thus the minimum beating period) between interfering states via 
\begin{equation}
\Delta=\sigma_{\omega}+\frac{1}{\sigma_T},
\end{equation}
where $\Delta$ is the maximum spacing between interfering states. Note that these resolutions always satisfy $\sigma_\omega\sigma_T\geq 1$ \cite{Sto94}. One can thus begin by applying a nearly-pure frequency-gating to eliminate interferences and then gradually widen the gating parameters in the frequency-domain to reveal the density of transitions at various parts of the spectra. A clear demonstration is the prominently-beating transitions with period $\sim 120$ fs at $\sim 9$ eV which is visible (to varying degrees) in all three 2D figures. An optimal balance for observing this interference is found in Fig. \ref{fig:2D}a in which the beating period (and thus the state seperation) can be well-approximated by visual inspection. In Fig. \ref{fig:2D}b $\Delta$ is so large that the interference of interest is masked by the myriad of other possible interferences while in Fig. \ref{fig:2D}c, $\Delta$ is small enough to almost entirely suppress the $\sim 9$ eV beating. The coherent DFG signal coincides with SRIF in the regime $\bar{\omega}>5$ eV and vanishes otherwise. It is thus, not plotted separately.

\begin{figure}[t]
\includegraphics[width=.35\textwidth]{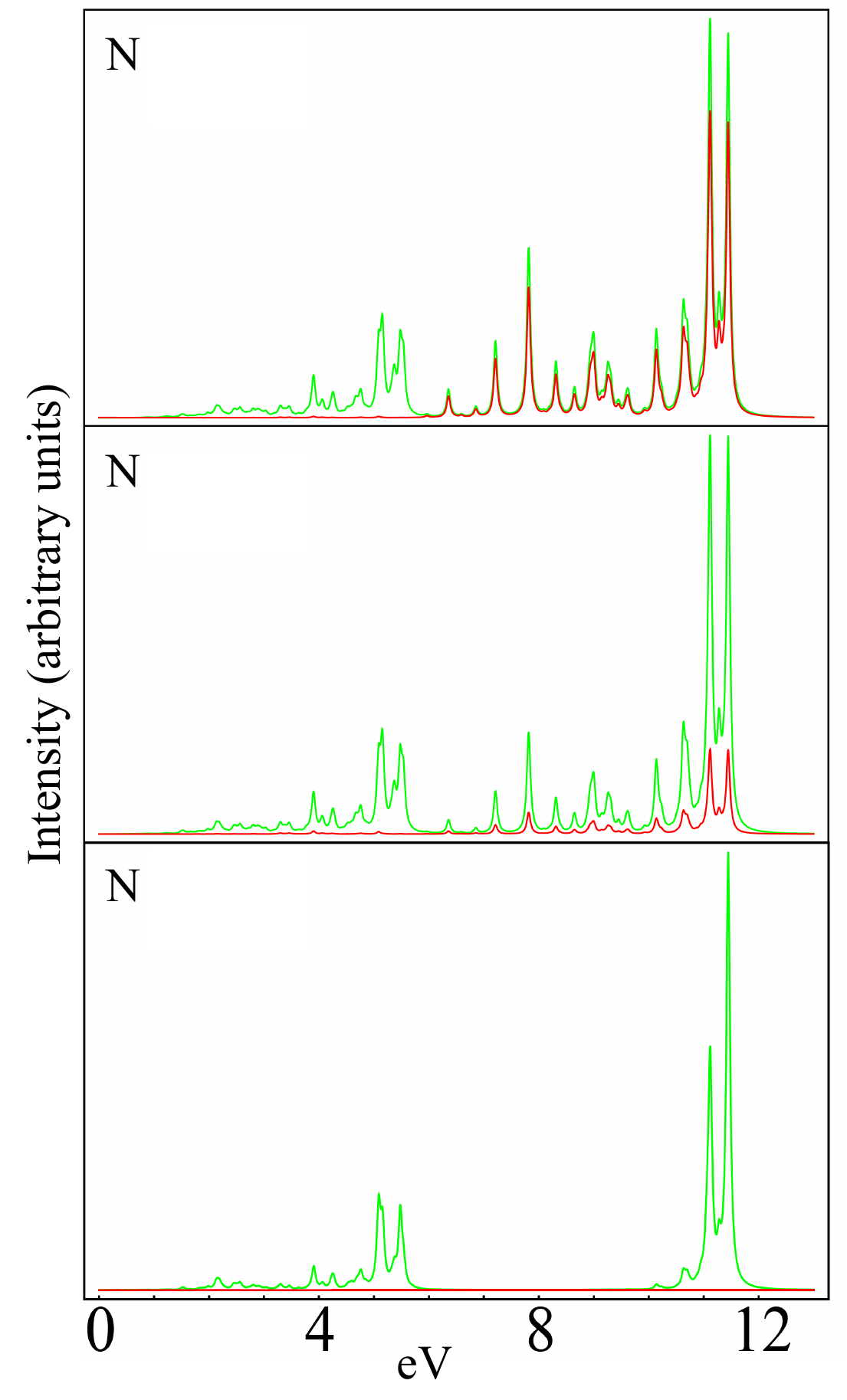}
\caption{(Color online) Frequency-resolved SRIF - green (light gray) line and DFG - red (dark gray) line following a stimulated Raman excitation by two pulses of varying width.  In all three panels, the first pulse is tuned to the transition from the bottom of the nitrogen core band to the ground state ($\sim$ 388 eV) and has a width of 1.36 eV.  The second pulse is tuned to the transition from the bottom of the nitrogen core band to the top of the valence band ($\sim$ 374 eV).  The second pulse is gradually broadened to allow for more valence excitations prior to the NLS $\sim$1.36 eV (bottom), $\sim$3eV (middle), and $\sim$6 eV (top). $\Gamma\sim$0.04 eV is used for the gating bandwidth.}\label{fig:2pulse}
\end{figure}

We note, that the choice $\Gamma=0.04$ eV in Figs. \ref{fig:cohinc}, \ref{fig:timedom} and \ref{fig:2pulse} represents an experimental resolution. The lifetime of the valence states, are of the order of ns and will give broadening $\Gamma\sim\mu$ eV. The detection parameters are chosen to capture the specific beating signal between various transitions that occur on much shorter time scale of tens of fs. The low quantum yield at short times poses a challenge for the detection.

\section{Discussion}

Thanks to the $\sim N^2$ versus $\sim N$-scaling it is not hard to separate the coherent signal since it is directed and generally dominates the incoherent and hyper-Rayleigh coherent signals.  However, it is not generally as simple to distinguish between the incoherent and hyper-Rayleigh contribution to the coherent signal since both are roughly isotropic and $\sim N$-scaling. In our model of cysteine, the valence manifold bandwidth is slightly larger than the lowest valence excitation and so the incoherent and coherent signals can be easily distinguished in the frequency domain, as shown in Section IV.

When the valence state manifold bandwidth is comparable to the band gap, both inter- and intra-band transitions  overlap spectrally and one needs to identify another procedure for separating the SRIF and DFG contributions. A single X-ray pulse cannot accomplish this goal since any modification of its bandwidth will affect both the absorption and the emission profile. However, this can be achieved  by using two X-ray  pulses with wave vectors $\mathbf{k}_1$ and $-\mathbf{k}_2$ for the system preparation via stimulated Raman excitation.  Both pulses must have narrow bandwidth compared to the $g$ manifold in order to excite states selectively. The first pulse is tuned to the bottom of the core-excitation manifold.  The second pulse is tuned to the transition from the bottom of the core-excitation manifold to the top of the valence band.  This combination of pulse parameters can selectively excite populations and coherences of the high-energy valence states. Since the excitation is weak, the square of the coherences ground and excited states are comparable to the valence populations but the coherences between valence states are lower order. Thus, the coherences between valence states created by this two-pulse process do not contribute significantly to the signal.  The overall signal will then be dominated by the incoherent contribution, in particular the transitions from high energy valence states to lower energy valence states and the ground state.  Increasing the bandwidth of pulse $-\mathbf{k}_2$ will result in the excitation of larger number of valence states which gives rise to a small coherent contribution as well as more transitions to appear in the incoherent spectra.  Finally, as the second pulse is broadened so that it excites the entire valence band, the incoherent and coherent signals again overlap in the regime of transitions to the ground state. This behavior is illustrated in Fig. \ref{fig:2pulse}.

In summary, by working in the joint field-matter space we developed QED expressions for the  NLS signal and show that may not be solely expressed in terms of the macroscopic polarization as suggested by the semiclassical approach. We find an isotropic hyper-Rayleigh and directional coherent component that can be described by the Larmor formula and an incoherent component that governs the fluorescence.  The microscopic calculation reveals interferences of the quantum matter pathways. 

\begin{acknowledgments}
The support of the Chemical Sciences, Geosciences, andBiosciences division, Office of Basic Energy Sciences, Office of Science, U.S. Department of Energy is gratefully acknowledged.  We also gratefully acknowledge the support of the National Science Foundation (Grant CHE-1058791), and the National Institutes of health (Grant GM-59230). 
\end{acknowledgments}

\appendix

\section{Microscopic Calculation of Spontaneous Light Emission}\label{app:mic}

Simultaneously time-and-frequency gated spontaneous light emission signals may be described by the spectral and temporal overlap of a bare signal and a detector spectrogram (Eq. (\ref{eq:S011})).
The detector spectrogram $W_D$ is an ordinary function of the gating time and frequency parameters which are characterized by standard deviations of the time and frequency gating $\sigma_T$ and $\sigma_{\omega}$, respectively. The structure of $W_D$ guarantees that these always satisfy the Fourier uncertainty $\sigma_\omega\sigma_T\geq 1$. The bare signal contains all of the relevant information about the molecules.

 In order to maintain the bookkeeping of all interactions and develop a perturbative expansion for signals we adopt superoperator notation. With each ordinary operator $O$ we associate a pair of superoperators \cite{Har08} ``left'' $\hat{O}_LX=OX$, ``right'' $\hat{O}_RX=XO$, and the combination $\hat{O}_-=\hat{O}_L-\hat{O}_R$. The bare spectrogram $W_B$ in the gated photon counting signal (\ref{eq:S011})  is given in terms of superoperators as:
\begin{align}\label{eq:WE01}
&W_B(t',\omega')=\int_0^{\infty}d\tau e^{-i\omega'\tau}\times\notag\\
& \langle \mathcal{T} \hat{\mathbf{E}}_{R}^{\dagger}(t'+\tau/2,\mathbf{r}_D)\hat{\mathbf{E}}_{L}(t'-\tau/2,\mathbf{r}_D)e^{-\frac{i}{\hbar}\int_{-\infty}^{\infty}\hat{H}_-'(T)dT}\rangle.
\end{align}
The positive frequency part of the electric field operator is given by
\begin{equation}\label{eq:Ejs}
\hat{\mathbf{E}}(t,\mathbf{r})=\sum_{\mathbf{k}_s,\mu}\left(\frac{2\pi\hbar \omega_s}{\Omega}\right)^{1/2}\epsilon^{(\mu)}(\mathbf{k}_s)\hat{a}_{\mathbf{k}_s}e^{-i\omega_st+i\mathbf{k}_s\cdot\mathbf{r}},
\end{equation}
where $\epsilon^{(\mu)}(\mathbf{k})$ is the unit electric polarization vector of mode $(\mathbf{k}_s,\mu)$, $\mu$ being the index of polarization, $\omega_s=c|\mathbf{k}_s|$, $c$ is speed of light, $\Omega$ is quantization volume. The polarization degrees of freedom are necessary to restrict the density of radiation modes and recover the $\lambda^{-4}$ dependence typical for Rayleigh scattering. The Hamiltonian superoperator in the interaction picture under the rotating-wave approximation (RWA) is given by:
\begin{equation}\label{eq:Hq}
 \hat{H}'_{q}(t)=\int d\mathbf{r}\hat{\mathbf{E}}_{q }^{\dagger}(t,\mathbf{r})\hat{\mathbf{V}}_{q}(t,\mathbf{r})+H.c,\quad q=L,R,
 \end{equation}
where $\mathbf{V}(t,\mathbf{r})=\sum_{\alpha}\mathbf{V}^{\alpha}(t)\delta(\mathbf{r}-\mathbf{r}_{\alpha})$ is a matter operator representing the lowering (exciton annihilation) part of the dipole coupling  and $\alpha$ runs over molecules in the sample located at $\mathbf{r}_{\alpha}$, $\mathbf{r}_{\beta}$. The operator $\mathcal{T}$ maintains positive time ordering of superoperators, and is a key bookkeeping device. It is defined as follows: 
 \begin{align}
\mathcal{T}\hat{E}_{q}(t_1)\hat{E}_{q'}(t_2)&=\theta(t_1-t_2)\hat{E}_{q}(t_1)\hat{E}_{q'}(t_2)\notag\\
&+\theta(t_2-t_1)\hat{E}_{q'}(t_2)\hat{E}_{q}(t_1),
\end{align}
where $\theta(t)$ is the Heaviside step function. Note that the interaction Hamiltonian (\ref{eq:Hq}) corresponds to the interaction of the matter with the spontaneously emitted photon between two electronic states.  This energy is in VUV range and is not related to the preparation of the system that involves an X-ray Raman process with high energy photons. Therefore, the dipole approximation for field-matter interaction is justified.

We shall calculate the time dependent bare signal (\ref{eq:WE01}) for a collection of molecules $\alpha$, $\beta$ in the interaction picture. We consider a signal governed by the spatial phase factor $e^{i\mathbf{k}_n\cdot \mathbf{r}}$, where $\mathbf{k}_n$ is a combination of incoming wave vectors characteristic to the desired signal. This phase factor represents a paraxial approximation, such that multiple frequency components propagate with the same wave vector. The bare signal represented by the closed path time-loop diagram shown in Fig. \ref{fig:det}a,b. We first  expand the time ordered exponent in Eq.(\ref{eq:WE01}) to second order - one interaction with the bra and one with the ket - and factorize the detected field and matter correlation functions
\begin{widetext}
\begin{align}\label{eq:fac1}
&\frac{1}{\hbar^2}\sum_{\alpha,\beta}^{N}\sum_{\nu,\nu'}\int_{-\infty}^{t'+\tau/2}dt_1\int_{-\infty}^{t'-\tau/2}dt_1'\langle \mathcal{T}\hat{V}^{\alpha}_{\nu L}(t_1')\hat{V}_{\nu'R}^{\beta\dagger}(t_1)\rangle\langle \mathcal{T}\hat{E}_{\mu R}^{(s)\dagger}(t'+\tau/2,\mathbf{r}_D)\hat{E}_{\mu L}^{(s')}(t'-\tau/2,\mathbf{r}_D)\hat{E}_{\nu L}^{(s')\dagger}(t_1',\mathbf{r}_{\alpha})\hat{E}_{\nu'R}^{(s)}(t_1,\mathbf{r}_{\beta})\rangle.
\end{align}
Since $\hat{E}^{(s)}$ is initially in the vacuum state, the field correlation function factorizes using (\ref{eq:Ejs}) and the bare signal (\ref{eq:WE01}) is given by 

\begin{align}\label{eq:WE02}
&W_B(t',\omega')=\left(\frac{2\pi}{\Omega}\right)^2\sum_{\mathbf{k}_s,\mathbf{k}_{s'}}\int_0^{\infty}d\tau e^{-i\omega'\tau}\int_{-\infty}^{t'+\tau/2}dt_1\int_{-\infty}^{t'-\tau/2}dt_1'\omega_s\omega_{s'}e^{i\omega_s(t'+\tau/2-t_1)-i\omega_{s'}(t'-\tau/2-t_1')}\notag\\
&\times\sum_{\alpha,\beta}^{N}e^{i(\mathbf{k}_s-\mathbf{k}_n)\cdot\mathbf{R}_{\beta}-i(\mathbf{k}_{s'}-\mathbf{k}_n)\cdot\mathbf{R}_{\alpha}}\sum_{\mu,\nu,\nu'}\epsilon^{(\nu)}(\mathbf{k}_s)\epsilon^{(\mu)}(\mathbf{k}_s)\epsilon^{(\nu')}(\mathbf{k}_{s'})\epsilon^{(\mu)}(\mathbf{k}_{s'})\langle \mathcal{T}\hat{V}^{\alpha}_{\nu L}(t_1')\hat{V}_{\nu'R}^{\beta\dagger}(t_1)e^{-\frac{i}{\hbar}\int_{-\infty}^{\infty}\hat{H}_-'(T)dT}\rangle,
\end{align}
where  $\mathbf{R}_{\alpha}=\mathbf{r}_{\alpha}-\mathbf{r}_D$ and $\mathbf{r}_{D}$ is the position of the detector. Eq. (\ref{eq:WE02}) contains explicitly multiple pairs of radiation modes $\mathbf{k}_s$ and $\mathbf{k}_{s'}$ and acts in the joint field plus matter space.  It takes into account all field matter interactions that lead to the emission of the detected field modes.

\section{Superoperator representation of coherent and incoherent signals}\label{app:super}

For the single-molecule (incoherent) signal $\mathbf{r}_{\alpha}=\mathbf{r}_{\beta}$ and thus, the signal reads

\begin{align}\label{eq:WEinc}
W_{Binc}(t',\omega')=\left(\frac{2\pi}{\Omega}\right)^2&\sum_{\mathbf{k}_s,\mathbf{k}_s'}\int_0^{\infty}d\tau e^{-i\omega'\tau}\sum_{\alpha}^Ne^{i(\mathbf{k}_s-\mathbf{k}_{s'})\cdot\mathbf{R}_{\alpha}}\int_{-\infty}^{t'+\tau/2}dt_1\int_{-\infty}^{t'-\tau/2}dt_1'\omega_s\omega_s'e^{i\omega_s(t'+\tau/2-t_1)-i\omega_s'(t'-\tau/2-t_1')}\notag\\
\times&\sum_{\mu,\nu,\nu'}\epsilon^{(\nu)}(\mathbf{k}_s)\epsilon^{(\mu)}(\mathbf{k}_s)\epsilon^{(\nu')}(\mathbf{k}_{s'})\epsilon^{(\mu)}(\mathbf{k}_{s'})\langle \mathcal{T}\hat{V}_{\nu L}(t_1')\hat{V}_{\nu' R}^{\dagger}(t_1)e^{-\frac{i}{\hbar}\int_{-\infty}^{\infty}\hat{H}_-'(T)dT}\rangle_{\alpha},
\end{align}

In the absence of dissipation (unitary evolution) we can further factorize the matter correlation function as 
\begin{align}\label{eq:Vfac}
&\langle \mathcal{T}\hat{V}_{\nu L}(t_1')\hat{V}_{\nu' R}^{\dagger}(t_1)\rangle=\sum_{a,b}\rho_{aa}\langle\langle aa|\hat{V}_{\nu' R}^{\dagger}(t_1)|ab\rangle\rangle\langle\langle ba| \hat{V}_{\nu L}(t_1')  |aa\rangle\rangle=\notag\\
&\sum_{a,b}\rho_{aa}\langle\langle b(t'+\tau/2)a| \hat{V}_{\nu L}(t_1')\mathcal{T}_+\exp\left[-\frac{i}{\hbar}\int_{-\infty}^{t_1'}\hat{H}_L'(T)dT\right]|aa\rangle\rangle\langle\langle aa|\hat{V}_{\nu' R}^{\dagger}(t_1)\mathcal{T}_-\exp\left[\frac{i}{\hbar}\int_{-\infty}^{t_1}\hat{H}_R'(T)dT\right]|b(t'+\tau/2)a\rangle\rangle,
\end{align}
where we denote $\langle\langle ba|\hat{O}|aa\rangle\rangle\equiv\text{Tr}[|a\rangle\langle b| \hat{O}|a\rangle\langle a|]$ and $b$ is the final state of the system. We assume that  system is initially in the pure state $a$ described by equilibrium density matrix $\rho_{aa}$. We next define the transition amplitude 
\begin{align}\label{eq:Teg}
\tilde{T}_{ba}^{(\alpha,\mu)}(t)=-&i\sum_{\mathbf{k}_s,\nu}\epsilon^{(\nu)}(\mathbf{k}_s)\epsilon^{(\mu)}(\mathbf{k}_s)\frac{2\pi\omega_s}{\Omega}\int_{-\infty}^tdt_1'e^{-i\omega_s(t-t_1')-i\omega_{ab}t-i(\mathbf{k}_s-\mathbf{k}_{\lbrace n\rbrace})\cdot\mathbf{R}_{\alpha}}\notag\\
&\times\langle\langle b(t)a|\hat{V}_{\nu L}(t_1')\mathcal{T}\exp\left(-\frac{i}{\hbar}\int_{-\infty}^{t_1'}\hat{H}_L'(T)dT\right)|aa\rangle\rangle_\alpha.
\end{align}
 Since all interactions are from the left ($L$), we can also write the transition amplitude using ordinary operators in Hilbert space
\begin{align}
\tilde{T}_{ba}^{(\alpha,\mu)}(t)=-&i\sum_{\mathbf{k}_s,\nu}\epsilon^{(\nu)}(\mathbf{k}_s)\epsilon^{(\mu)}(\mathbf{k}_s)\frac{2\pi\omega_s}{\Omega}\int_{-\infty}^tdt_1'e^{-i\omega_s(t-t_1')-i\omega_{ab}t-i(\mathbf{k}_s-\mathbf{k}_{\lbrace n\rbrace})\cdot\mathbf{R}_{\alpha}}\notag\\
&\times\langle b(t)|V_\nu(t_1')\mathcal{T}\exp\left(-\frac{i}{\hbar}\int_{-\infty}^{t_1'}H'(T)dT\right)|a\rangle_\alpha
\end{align}
\end{widetext}
This gives for the bare signal (\ref{eq:WEinc})
\begin{align}\label{eq:WE102}
W_{Binc}(t',\omega')&=\sum_{a,b,\mu}\sum_{\alpha=1}^N\rho_{aa}\int_0^{\infty}d\tau e^{-i\omega'\tau}\notag\\
&\times \tilde{T}_{ba}^{(\alpha,\mu)}(t'-\tau/2)\tilde{T}_{ab}^{(\alpha,\mu)*}(t'+\tau/2).
\end{align}
In the limit of pure time-resolved measurement (\ref{eq:St}) signal (\ref{eq:WE102}) transforms into
\begin{equation}
W_{Binc}(\bar{t})=\sum_{a,b,\mu}\sum_{\alpha=1}^N\rho_{aa}|\tilde{T}_{ba}^{(\alpha,\mu)}(\bar{t})|^2,
\end{equation}
Similarly the pure frequency-resolved signal (\ref{eq:Sw}) signal (\ref{eq:WE102}) yields
\begin{equation}
W_{Binc}(\bar{\omega})=\sum_{a,b,\mu}\sum_{\alpha=1}^N\rho_{aa}|\tilde{T}_{ba}^{(\alpha,\mu)}(\bar{\omega})|^2,
\end{equation}
where $\tilde{T}_{ba}^{(\alpha,\mu)}(\omega)=\int_{-\infty}^{\infty}dt\tilde{T}_{ba}^{(\alpha,\mu)}(t)e^{i\omega t}$. The spatial phase factors in Eq. (\ref{eq:WE02}) indicates that the incoherent emission occurs in all directions and is independent of the wave vector of the incoming pulses as expected. A single summation over the molecule yields the linear scaling with respect to the number of molecules.

We now turn to coherent emission  contribution of $\alpha\neq\beta$. Since interactions with different molecules are not time-ordered, we may factorize the matter correlation in (\ref{eq:WE02})  $\langle \mathcal{T}\hat{V}^{\alpha}_L(t_1')\hat{V}_R^{\beta\dagger}(t_1)\rangle \to\langle \hat{V}^{\alpha}_L(t_1')\rangle \langle\hat{V}_R^{\beta\dagger}(t_1)\rangle$. Thus, the two-molecule (coherent) signal can be separated into a long-range and a short-range component (as described in \cite{Ros10}). Using Eqs. (\ref{eq:Vfac}) - (\ref{eq:Teg}) the coherent part of the signal (\ref{eq:WE02}) reads
\begin{align}\label{eq:WE101}
&W_{Bcoh}(t',\omega')=\notag\\
&\sum_{\mu}\sum_{\alpha=1}^N\sum_{\beta=1}^{N-1}\int_0^{\infty}d\tau e^{-i\omega'\tau}\tilde{P}^{(\alpha,\mu)}(t'-\tau/2)\tilde{P}^{(\beta,\mu)}(t'+\tau/2),
\end{align}
where 
\begin{widetext}
\begin{align}\label{eq:Peg1}
\tilde{P}^{(\alpha,\mu)}(t)=-&i\sum_{\mathbf{k}_s,\nu}\epsilon^{(\nu)}(\mathbf{k}_s)\epsilon^{(\mu)}(\mathbf{k}_s)\frac{2\pi\omega_s}{\Omega}\int_{-\infty}^tdt_1'e^{-i\omega_s(t-t_1')-i\omega_{ab}t-i(\mathbf{k}_s-\mathbf{k}_{\lbrace n\rbrace})\cdot\mathbf{R}_{\alpha}}\notag\\
&\times\sum_{a,b}\rho_{aa}\langle\langle bb|\hat{V}_{\nu L}(t_1')\mathcal{T}\exp\left(-\frac{i}{\hbar}\int_{-\infty}^{t_1'}\hat{H}_L'(T)dT\right)|aa\rangle\rangle_\alpha.
\end{align}
\end{widetext}
Here $\rho_{aa}$ represents the initial equilibrium density matrix. Note that in contrast with Eq. (\ref{eq:Teg}) which represents the transition amplitude and thus can be recast in Hilbert space, (\ref{eq:Peg1}) is related to polarization and therefore can be written in Liouville space only. In the limit of pure time-resolved measurement (\ref{eq:St}) signal (\ref{eq:WE101}) transforms into
\begin{equation}
W_{Bcoh}(\bar{t})=\sum_{\mu}\sum_{\alpha=1}^N\sum_{\beta=1}^{N-1}\tilde{P}^{(\alpha,\mu)}(\bar{t})\tilde{P}^{(\beta,\mu)}(\bar{t}),
\end{equation}
Similarly the pure frequency-resolved signal (\ref{eq:Sw}) signal (\ref{eq:WE101}) yields
\begin{equation}
W_{Bcoh}(\bar{\omega})=\sum_{\mu}\sum_{\alpha=1}^N\sum_{\beta=1}^{N-1}\tilde{P}^{(\alpha,\mu)}(\bar{\omega})\tilde{P}^{(\beta,\mu)}(\bar{\omega}),
\end{equation}

\section{Microscopic derivation of the Larmor formula}\label{app:Lar}

In order to calculate the signal (\ref{eq:WE02}) we start with amplitude expression (\ref{eq:Teg}). We first evaluate the summation over the modes. In the continuum limit
\begin{equation}
\frac{1}{\Omega}\sum_{\mathbf{k}_s}=\frac{1}{(2\pi)^3}\int d\mathbf{k}_s.
\end{equation}
Recalling that $\hat{\mathbf{k}}_s$ and polarization vectors $\epsilon^{(\nu)}(\mathbf{k}_s)$, $\nu=1,2$ form a set of mutually perpendicular unit vectors it follows that
\begin{equation}
\epsilon^{(\nu)}(\mathbf{k}_s)\epsilon^{(\mu)}(\mathbf{k}_s)=\delta_{\mu\nu}-\hat{k}_{\mu}\hat{k}_{\nu}.
\end{equation}
We further evaluate the momentum integral using $d\mathbf{k}_s=k_s^2dk_sd\Omega$. The angular integration yields
\begin{align}
\frac{1}{4\pi}&\int d\Omega(\delta_{\mu\nu}-\hat{k}_\mu\hat{k}_\nu)e^{i\mathbf{k}_s\cdot \mathbf{R}}\notag\\
&=\frac{1}{k_s^3}(-\nabla^2\delta_{\mu\nu}+\nabla_\mu\cdot\nabla_\nu)\frac{\sin(k_sR)}{R}.
\end{align}
Eq. (\ref{eq:Teg}) then yields
\begin{align}
&\tilde{T}_{ba}^{(\alpha,\mu)}(t)=\frac{e^{i\mathbf{k}_n\cdot\mathbf{R}_\alpha}}{\pi}\sum_\nu\int\frac{d\omega}{2\pi}T_{ba}^{(\alpha,\nu)}(\omega)e^{-i\omega t}\notag\\
&\times\int d\omega_s(-\nabla^2\delta_{\mu\nu}+\nabla_\mu\cdot\nabla_\nu)\frac{\sin(\omega_sR_\alpha/c)}{R_\alpha}\frac{1}{\omega-\omega_s+i\epsilon},
\end{align}
where we introduced a matter transition amplitude $T_{ba}(\omega)$ that connects initial state $a$ with final state $b$:  $T_{ba}^{(\alpha,\nu)}(\omega)=\int_{-\infty}^{\infty}dte^{i\omega t}T_{ba}^{(\alpha,\nu)}(t)$ and $T_{ba}^{\alpha,\nu)}(t)$ is given by Eq. (\ref{eq:Ppr}). We then note that
\begin{align}
\int d\omega_s\frac{\sin(\omega_s R_\alpha/c)}{\omega_s-\omega-i\epsilon}=\pi e^{i\omega R_\alpha/c}.
\end{align}
Taking into account that \cite{Salam}
\begin{align}
&(-\nabla^2\delta_{\mu\nu}+\nabla_\mu\cdot\nabla_\nu)\frac{e^{ikR}}{R}=\notag\\
&\frac{1}{R^3}[(\delta_{\mu\nu}-3\hat{R}_\mu\hat{R}_\nu)(ikR-1)+(\delta_{\mu\nu}-\hat{R}_\mu\hat{R}_\nu)k^2R^2]e^{ikR}
\end{align}
and assuming the random orientation of the molecules: $\hat{R}_\mu\hat{R}_\nu=\frac{1}{3}\delta_{\mu\nu}$ we obtain
\begin{align}
\tilde{T}_{ba}^{(\alpha,\mu)}(t)=-\frac{2}{3c^2}\frac{e^{i\mathbf{k}_n\cdot\mathbf{R}_\alpha}}{R_{\alpha}}\partial_t^2T_{ba}^{(\mu,\alpha)}(t-R_\alpha/c).
\end{align}
Taking into account (\ref{eq:WE102}) we therefore obtain  (\ref{eq:Winc1}). Similarly we derive for coherent signal
\begin{align}
\tilde{P}^{(\alpha,\mu)}(t)=-\frac{2}{3c^2}\frac{e^{i\mathbf{k}_n\cdot\mathbf{R}_\alpha}}{R_{\alpha}}\partial_t^2P^{(\mu,\alpha)}(t-R_\alpha/c).
\end{align}
Thus, coherent signal Eq. (\ref{eq:WE101}) can be recast in the form of (\ref{eq:Wcoh1}).

\section{Semiclassical theory of emission detection from a radiating dipole}\label{app:sc}

According to classical electrodynamics, the electric field obeys the usual homogeneous wave equation derived from Maxwell
equations:
\begin{equation}\label{eq:wave}
\nabla \times\nabla\times \mathbf{E}(\mathbf{r},t)-\frac{1}{c^2}\frac{\partial^2\mathbf{E}(\mathbf{r},t)}{\partial t^2}=\frac{1}{\epsilon_0 c^2}\frac{\partial^2\tilde{\mathbf{P}}(\mathbf{r},t)}{\partial t^2}
\end{equation}
Here $\tilde{\mathbf{P}}$ is the macroscopic polarization and $\epsilon_0$ is the vacuum permittivity. Since Eq. (\ref{eq:wave}) is linear it applies also for a
quantum field, where the electric field $\mathbf{E}$ and $\tilde{\mathbf{P}}$ become operators. Restricting ourselves to the transverse part in far field $E=\mathbf{E}_{\perp}$, this equation can be recast in frequency domain as
\begin{equation}\label{eq:wave2}
\nabla^2E(\mathbf{r},\omega)-\frac{\omega^2}{c^2}E(\mathbf{r},\omega)=\frac{\omega^2}{\epsilon_0 c^2}\tilde{P}(\mathbf{r},\omega)
\end{equation}
The Green's function solution of Eq. (\ref{eq:wave2}) in infinite space for a single-point dipole molecule at $\mathbf{r}=0$
\begin{equation}
E(\mathbf{r},\omega)=-\frac{1}{\epsilon_0c^2}\omega^2\frac{e^{ik_0|\mathbf{r}|}}{4\pi|\mathbf{r}|}\tilde{P}(0,\omega).
\end{equation}
where $k_0=\omega/c$. Latter can be recast in the time domain
\begin{equation}
E(\mathbf{r},t)=\frac{1}{\epsilon_0 c^2}\frac{1}{4\pi|\mathbf{r}|}\partial_t^2\tilde{P}(0,t-|\mathbf{r}|/c)
\end{equation}
Neglecting retardation effects the electric field entering the detector at position $r_G$ is given
by 
\begin{equation}
E(r_G,t)=B\ddot{P}(t),
\end{equation}
where $B=(4\pi\epsilon_0c^2|r_G|)^{-1}$ and $P(t)\equiv \tilde{P}(0,t)$. We thus recover the Larmor formula for the time-resolved intensity of the signal
\begin{equation}
S(t)=B^2|\ddot{P}(t)|^2.
\end{equation}
Generally the signal will be affected by the detector geometry, depending if
an angle is used with a lens or if it enters in parallel. These
details will affect the prefactor $B$ and are not considered here.
The bare signal can be finally calculated as
\begin{align}\label{eq:WBP}
W_B(t',\omega')&=B^2\int_0^{\infty}d\tau e^{-i\omega'\tau}\ddot{P}^{*}(t'+\tau/2)\ddot{P}(t'-\tau/2).
\end{align}

%

\end{document}